\documentclass[runningheads]{llncs}
\usepackage{graphicx}
\usepackage{xcolor}
\usepackage{url}
\usepackage{wrapfig}
\usepackage{hyperref}

\begin{document}
\title{Passlab: A Password Security Tool \\ for the Blue Team}
%
%
\author{Saul Johnson\orcidID{0000-0001-9876-3775}}
\authorrunning{S. Johnson}
%
\institute{Teesside University, Middlesbrough, UK \\
\email{saul.johnson@tees.ac.uk}}
\maketitle              
\begin{abstract}
If we wish to compromise some password-protected system as an attacker (i.e. a member of the \textit{red team}), we have a large number of popular and actively-maintained tools to choose from in helping us to realise our goal. Password hash cracking hardware and software, online guessing tools, exploit frameworks, and a wealth of tools for helping us to perform reconnaissance on the target system are widely available. By comparison, if we wish to defend a password-protected system against such an attack (i.e. as a member of the \textit{blue team}), we have comparatively few tools to choose from. In this research abstract, we present our work to date on \textsc{Passlab}, a password security tool designed to help system administrators take advantage of formal methods in order to make sensible and evidence-based security decisions using a clean and intuitive user interface.
\end{abstract}
\section{Introduction}\label{sec:introduction}
When it comes to making decisions about the most appropriate password security policy for a system, the tendency has traditionally been to take a ``common sense'' approach. For example, it stands to reason intuitively that forcing a user to include at least one number in their password will make that password harder to guess and therefore more secure. Applying formal methods in order to quantify this increase in security, however, is seldom part of the decision-making process, leading to widely varying password security policies born from equally variable intuitions about which factors contribute to their security. Previous work finds that the password composition policy in place on a system has little to no correlation with the value of the assets it protects \cite{florencio2010where}. We expect that tightening legislation around data protection (in Europe in particular \cite{gdpr2016}) will encourage industry to invest in tools that offer the ability to make data-driven password security policy decisions.

The expected contribution is \textsc{Passlab}, an integrated environment that will allow system administrators without a background in formal methods to make informed password security decisions, formally reason about password composition policies and extract correct-by-construction software for enforcing them.

\section{Literature Review}\label{sec:literature}
We draw on work exploring the distribution of passwords by Malone and Maher \cite{malone2012investigating} and Wang et al. \cite{wang2017zipfs} in order to derive equations such as that in Figure~\ref{fig:passlab-eq} (see Section~\ref{subsec:lockout-policies}). The model of password composition policies presented by Blocki et al. in \cite{blocki2013optimizing} has shown promise thus far as a general low-level representation that allows us to encode password composition policies for a wide variety of password quality checking software. Work on attack-defence trees by Kordy et al. \cite{kordy2011foundations} offers us a way to allow system administrators without a formal methods background to model password guessing attacks and password composition policies to mitigate them in an intuitive and visual way. The Coq interactive theorem proving software \cite{bertot2013interactive} and its code extraction capabilities \cite{letouzey2008extraction} offer us a means to create usable, formally-verified password composition policy enforcement software.

\section{Progress to Date}\label{sec:progress}
In this section, we present progress to date on \textsc{Passlab} itself, and on work that forms the foundations of the formal methods it employs.

\subsection{Data-Informed Lockout Policies}\label{subsec:lockout-policies}
A \textit{lockout policy} is a restriction on the number of times a user can incorrectly enter their password before their account is locked down, requiring additional authentication via some other mechanism in order to reinstate their ability to log in. This offers strong protection against so-called \textit{online} password guessing attacks (i.e. attacks against the live service) but in turn introduces a denial-of-service vulnerability. If an attacker wants to prevent a user from accessing their account, they need only attempt to access it with the wrong password enough times that the lockout policy is triggered. This motivates us to search for formal methods to derive the maximum number of incorrect login attempts we can grant a user while guaranteeing that the probability of guessing attack success is kept below a specified threshold in the worst case. 

\begin{figure}
\centering
\includegraphics[width=0.85\textwidth]{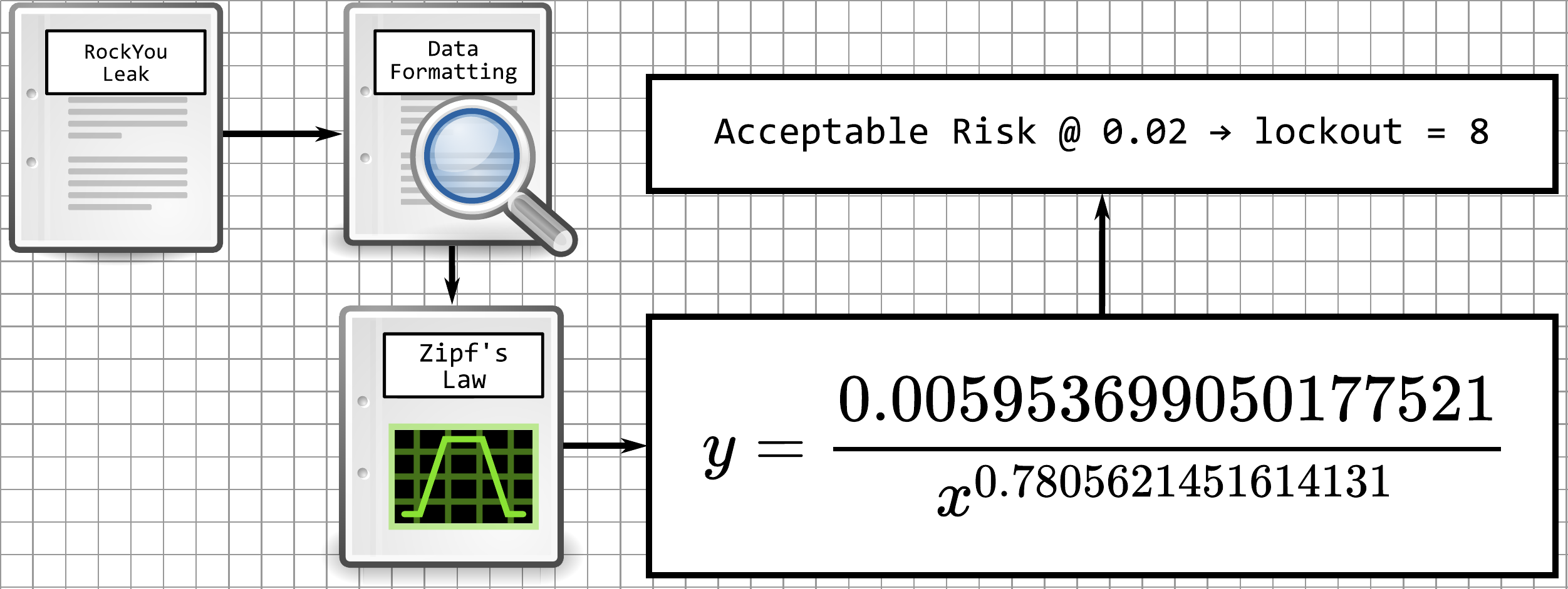}
\caption{A mock-up of the \textsc{Passlab} user interface. A raw data source node loads a raw password data dump (top left) which is then formatted (top centre) to convert it to a CSV file. After formatting, the data enters a Zipf model node (bottom centre) which computes a power-law equation to approximate guess success probability from password guessing order (bottom right).}
\vskip -1.5em
\label{fig:passlab-eq}
\end{figure}

Figure~\ref{fig:passlab-eq} shows a render of the \textsc{Passlab} user interface as it fits a power-law equation that maps the probability of a password being a correct guess ($x$) to its rank ($y$) in a large password data dump (in this case, the RockYou data set \cite{cubrilovich2009rockyou}). The software allows users to visually compose data analysis tasks such as that illustrated in Figure~\ref{fig:passlab-eq}. This draws on previous research, which finds that user-chosen passwords tend to follow Zipf's law in the general case \cite{malone2012investigating,wang2017zipfs}. That is, the frequency (and therefore probability) of a password is inversely proportional to its position in the data set, when ranked by frequency. In this case, it is possible to use this equation to calculate that, even if an attacker knew the most common 8 passwords on our system, if they selected an account at random and tried these 8 passwords they would have a probability of successfully gaining entry to that account of less than $0.02$.
\newpage
\subsection{Interactive Security Policy Building}\label{subsec:interactive-building}
\begin{wrapfigure}{L}{0.35\textwidth}
  \vspace{-2em}
  \centering
    \includegraphics[width=0.35\textwidth]{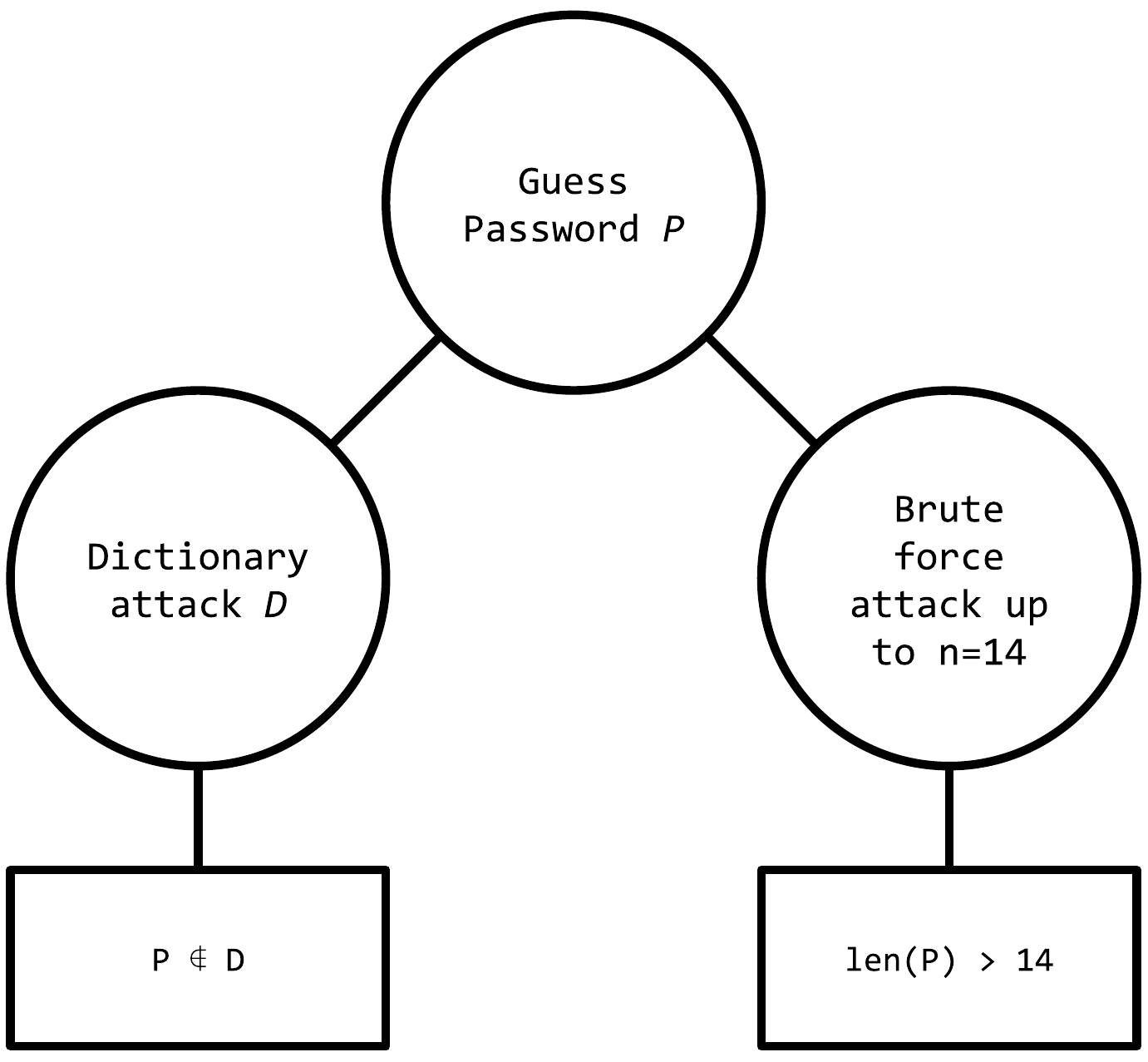}
  \caption{An abstract example of an attack-defence tree (ADTree) \cite{kordy2011foundations} from which a password composition policy might be synthesised.} \label{fig:passlab-adt}
  \vskip -0.5em
\end{wrapfigure}
\textsc{Passlab} will include an ``interactive security policy builder'' from within which a system administrator can model a password guessing attack and its mitigation measures as an attack-defence tree (ADTree) \cite{kordy2011foundations} and synthesise password composition policy enforcement software (see Section~\ref{subsec:c-by-c}). 
Figure~\ref{fig:passlab-adt} shows 
how such an ADTree might look, with policy-level mitigations for each mode employed by the password guessing attack in place. The attacker has a goal of guessing a password and, to try and achieve this, employs a bimodal guessing attack---a dictionary attack using dictionary $D$ and a brute-force attack of passwords of length up to 14. To mitigate this, we add defence nodes to ensure that the password is not contained in $D$ and that its length is greater than 14.

After the system administrator builds a security policy in this way, we would like them to be able to perform further reasoning from outside that environment. Work on this is ongoing in the form of \textit{Skeptic}, our Coq framework for reasoning about password composition policies \cite{skeptic19}, which has already shown promise in formally verifying that certain password composition policies confer immunity to certain attacks, for example. Creating software that combines analysis of large data sets with formal verification for usable security continues to be one of the main challenges in our research.

\subsection{Correct-by-Construction Enforcement Software}\label{subsec:c-by-c}
We have previous work \cite{ferreira2017certified} on creating certified password composition policy enforcement software, implemented from within the Coq proof assistant \cite{bertot2013interactive} and extracted to Haskell \cite{letouzey2008extraction}. The extracted Haskell is then compiled into a pluggable authentication module readily usable from a real Linux system. Building on this work, we created \textit{Serenity}~\cite{serenity19}, a DSL for building correct-by-construction password quality checking software that employs formal verification to ensure that password composition policies built using it are correct. A small user study has provided encouraging preliminary results indicating that users find it easier to express their desired password composition policy using the Serenity DSL than they do using more traditional password composition policy enforcement software.

%
%
\bibliographystyle{splncs04}
\bibliography{main}

\end{document}